\documentclass[a4paper,twocolumn,showpacs,superscriptaddress,prl]{revtex4}
\usepackage[english]{babel}
\usepackage{graphicx}
\usepackage{xspace}
\usepackage{amsmath,amssymb}
\usepackage{subfigure}
\DeclareMathAlphabet{\mathbfit}{OT1}{cmr}{bx}{it}

\begin{document}

\title{Room temperature tunneling magnetoresistance in magnetite
based junctions: Influence of tunneling barrier}

\author{D.~Reisinger}
\email{Daniel.Reisinger@wmi.badw.de}
\affiliation{Walther-Mei{\ss}ner-Institut, Bayerische Akademie der
Wissenschaften, Walther-Mei{\ss}ner Str.~8, 85748 Garching,
Germany}

\author{P.~Majewski}
\affiliation{Walther-Mei{\ss}ner-Institut, Bayerische Akademie der
Wissenschaften, Walther-Mei{\ss}ner Str.~8, 85748 Garching, Germany}

\author{M.~Opel}
\affiliation{Walther-Mei{\ss}ner-Institut, Bayerische Akademie der
Wissenschaften, Walther-Mei{\ss}ner Str.~8, 85748 Garching, Germany}

\author{L.~Alff}
\email{Lambert.Alff@wmi.badw.de}
\affiliation{Walther-Mei{\ss}ner-Institut, Bayerische Akademie der
Wissenschaften, Walther-Mei{\ss}ner Str.~8, 85748 Garching, Germany}

\author{R.~Gross}
\affiliation{Walther-Mei{\ss}ner-Institut, Bayerische Akademie der
Wissenschaften, Walther-Mei{\ss}ner Str.~8, 85748 Garching,
Germany}

\date{received March 24, 2004}
\pacs{75.70.-i, 85.75.-d}

\begin{abstract}

Magnetite (Fe$_3$O$_4$) based tunnel junctions with turret/mesa
structure have been investigated for different barrier materials
(SrTiO$_3$, NdGaO$_3$, MgO, SiO$_2$, and Al$_2$O$_{3-x}$).
Junctions with a Ni counter electrode and an aluminium oxide
barrier showed reproducibly a tunneling magnetoresistance (TMR)
effect at room temperature of up to 5\% with almost ideal
switching behavior. This number only partially reflects the
intrinsic high spin polarization of Fe$_3$O$_4$. It is
considerably decreased due to an additional series resistance
within the junction. Only SiO$_2$ and Al$_2$O$_{3-x}$ barriers
provide magnetically decoupled electrodes as necessary for sharp
switching. The observed decrease of the TMR effect as a function
of increasing temperature is due to a decrease in spin
polarization and an increase in spin-scattering in the barrier.
Among the oxide half-metals magnetite has the potential to enhance
the performance of TMR based devices.

\end{abstract}
\maketitle
\vspace{0.5cm}
\section{Introduction}

Magnetic random access memory (MRAM) devices based on magnetic
tunnel junctions (MTJ) using ferromagnetic metals and alloys with
limited spin polarization will be implemented in next generation
computer memory. Beyond purely storage usage, MTJs can at the same
time be used as programmable logic elements \cite{Ney:03}. While
the device preparation techniques using simple metals and alloys
are at least in principle well known, from the point of view of
magnetic properties half metallic materials certainly are superior
to classical ferromagnetic metals. Magnetite (Fe$_3$O$_4$) is an
interesting candidate, because it has been predicted to
be a half-metal even at room temperature due to its high ordering
temperature of 860\,K \cite{Zhang:91,Jeng:02}. Indeed,
spin-resolved photoelectron spectroscopy on magnetite thin films
has recently revealed a spin polarisation near the Fermi edge of
up to 80\% at room temperature \cite{Dedkov:02}. In contrast, in
MTJs only very small TMR effects have been observed
\cite{Li:98,Ghosh:98,Seneor:99,Zaag:00} with a maximum effect of
14\% \cite{Matsuda:02}. Therefore, it is important to understand
the behavior of magnetite at the electrode/barrier interfaces, the
influence of the tunneling barrier itself, and the magnetic
coupling through thin barriers.

In this paper we report measurements of MTJs based on epitaxial
magnetite thin films on MgO(001) single crystal substrates. As
tunnel barrier the five materials MgO, SrTiO$_3$, NdGaO$_3$,
SiO$_2$, and Al$_2$O$_{3-x}$ have been investigated. As
counter-electrode Ni was used with about 33\% spin polarisation as
known from tunneling measurements \cite{Moodera:99}. The whole
thin film structure was grown by pulsed laser deposition (PLD) and
electron beam evaporation in an ultra high vacuum system. The
tunnel junctions with areas ranging from $10\times 10\,\mu$m$^2$
to $20\times 40\,\mu$m$^2$ were fabricated by optical lithography
and ion beam etching. The magnetic properties, in particular the
coupling of the electrodes through the barrier, have been studied
by SQUID magnetometry.  The magnetotransport behavior of these
MTJs was measured as a function of temperature and applied
magnetic field. \vspace{-0.5cm}

\section{Experimental Techniques}

The thin films were fabricated in an ultra high vacuum (UHV) laser
ablation system with in-situ electron beam evaporation. The
magnetite thin films were grown epitaxially on MgO single crystal
substrates, mostly with an underlying epitaxial TiN buffer layer.
Details about the whole process including in-situ reflection high
energy electron diffraction (RHEED) and laser substrate heating
are described in \cite{Klein:99,Reisinger:03a,Reisinger:03b}. To
enhance the surface quality a telescope optics for the laser beam
has been developed, so that for each material the optimal energy
density in combination with an optimal focus could be easily
achieved. This was especially important for getting smooth
surfaces without droplets which may lead  to short circuits
through the tunnel barrier. The surface quality was probed in-situ
with an Omicron atomic force microscope, the crystal properties
and film thickness with a Bruker/AXS high resolution
X-ray-diffractometer, and the magnetic properties with a Quantum
Design superconducting quantum interference device (SQUID)
magnetometer. The magneto-transport measurements were performed in
an Oxford cryostat system with variable temperature insert and a
10\,Tesla superconducting magnet. The magnetite thin films
achieved nearly the same properties as single crystal bulk
material with respect to magnetization on the Verwey-transition
behavior \cite{Reisinger:04}. Such high-quality thin films seem to
be one of the key prerequisites in order to obtain TMR devices
with desired performance.
\newpage

\begin{figure}[t]
 \centering
 \subfigure[\label{tmr-bruecke}]{\includegraphics[width=0.75\columnwidth,trim=0 0 0 0]{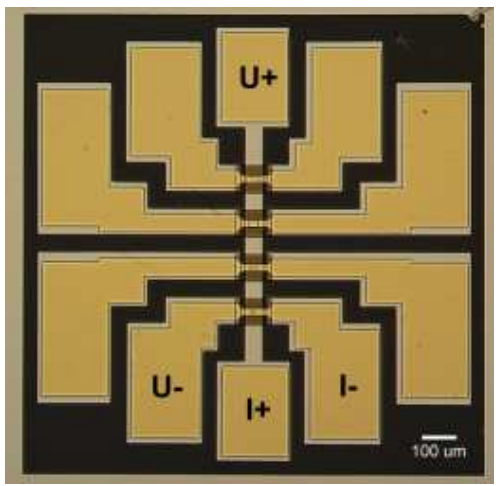}}
 \\
 \vspace{-0.5cm}
 \subfigure[\label{tmr-contact}]{\includegraphics[width=0.75\columnwidth, trim=0 0 0 0]{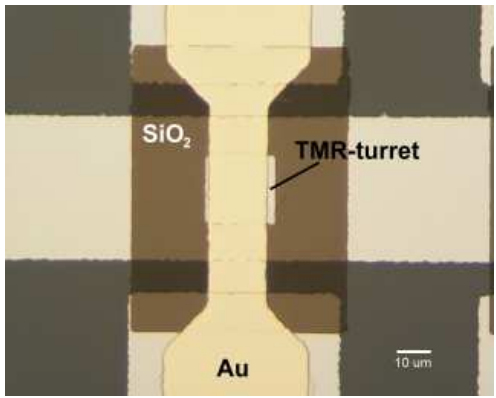}}
 \caption{Optical microscope images of a completed TMR-bridge (a) with four
 TMR-contacts (called turrets), and a zoom-view (rotated by 90$^\circ$) of a single turret with a size of
 $20\times20\,\mu$m$^2$ (b).}
 \label{fig1}
 \vspace{-0.5cm}
\end{figure}

On top of the 40~to~50\,nm thin magnetite films, 2~to~7\,nm thin
layers serving as tunnel barriers are deposited. Five different
materials were used: MgO, SrTiO$_3$ and NdGaO$_3$ have been
fabricated by PLD from stoichiometric targets, SiO$_2$ and Al
layers were deposited by electron beam evaporation in UHV. During
PLD growth the Ar-atmosphere was 10$^{-3}$~mbar, and the substrate
temperature was 330$^{\circ}$C. The only material growing
epitaxially on Fe$_3$O$_4$ is MgO as confirmed by the
corresponding RHEED pattern. Epitaxial growth of MgO is also
achieved at room temperature for a pure oxygen atmosphere
(10$^{-3}$~mbar). The SiO$_2$ and Al layers were deposited without
breaking the UHV after the magnetite deposition. Both materials
are grown at room temperature. Afterwards the Al was oxidized in
pure oxygen to form the insulating Al$_2$O$_{3-x}$. SrTiO$_3$,
NdGaO$_3$, SiO$_2$, and Al$_2$O$_{3-x}$ all form amorphous layers
on magnetite due to the large crystal lattice mismatch. In a
further step, a 40\,nm Ni layer was deposited in situ as second
magnetic electrode by electron beam evaporation in UHV at room
temperature.

Starting from these multilayers, tunnel junctions have been
fabricated using optical lithography and ion beam etching. The
complex production process consists of roughly 35 critical steps.
After careful optimization a yield of more than 50\% of
reproducible TMR-contacts was achieved corresponding to a yield of
98\% for each step. This high reproducibility is an important
prerequisite for a reliable analysis of the TMR-effect. As show in
Fig.~\ref{tmr-bruecke} one measurement bridge contains four tunnel
junction in form of small \glq turrets\grq\ with areas ranging
from $10\times 10\,\mu$m$^2$ to $20\times 40\,\mu$m$^2$. On a
$5\,\text{mm}\times5\,\text{mm}$ substrate four such structures
can be patterend. Fig.~\ref{tmr-contact} shows a zoom-view of one
small TMR-turret. The measurement current flows from the main
bottom electrode path, up through the turret with the magnetic
electrode layers and the tunnel barrier perpendicular to the
sample surface, and then back through the upper narrow-waisted
gold electrode.


\section{Experimental Results and Discussion}
\subsection{Magnetization}

The magnetization as a function of temperature and applied
magnetic field of the multilayer samples has been measured before
any lithographic step i.e.~the whole chip was measured. To
investigate the influence of the lithographic patterning process,
one sample has been patterned into a
$3\,\text{mm}\times3\,\text{mm}$ square. The magnetization
behavior of the sample before and after patterning was almost
identical. This means that the magnetization data as measured for
the whole chip is valid approximately also for the patterned
TMR-contacts.
\begin{figure}[h!]
 \centering
  \includegraphics[width=0.9\columnwidth, trim=10 10 10 10]{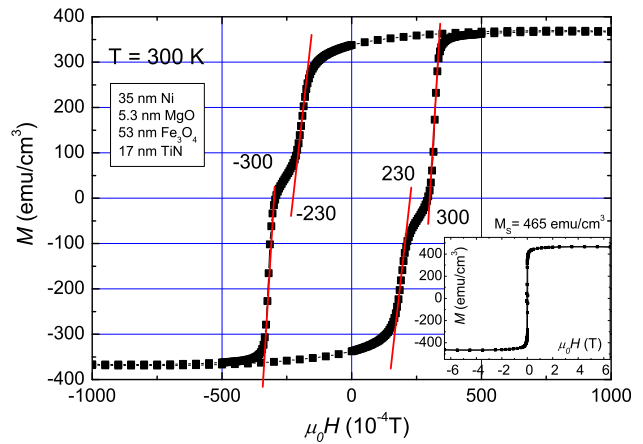}
 \caption{Magnetization behavior of a TiN/Fe$_3$O$_4$/MgO/Ni multilayer.
 Inset: Same measurement up to 6\,T.}
 \label{fig2}
\end{figure}

First, we consider the saturation magnetization $M_{\text{S}}$.
The inset of Fig.~\ref{fig2} shows a magnetization measurement of
a TiN/Fe$_3$O$_4$/MgO/Ni multilayer up to above 6\,T. For single
reference layers of magnetite and Ni we have measured
$M_{\text{S}}\approx453\,\text{emu/cm$^3$}$
resp.~$M_{\text{S}}\approx495\,\text{emu/cm$^3$}$ at room temperature. Knowing the
film thicknesses within the multilayer consisting of 53\,nm
magnetite, 5.3\,nm MgO, and 35\,nm Ni on top of a 17\,nm TiN
buffer layer, a saturation magnetization of
$470\,\text{emu/cm$^3$}$ is expected which is in excellent
agreement with the measured value of $465\,\text{emu/cm$^3$}$.
Together with the clear step-like shape of the hysteretic
$M(H)$-curve this is an indication of nearly ideal ferromagnetic
order in the electrode layers of the multilayer structures. This
is important with respect to the formation of magnetic domains or
glass-like behavior which both is unfavorable for the device
performance. Note, that $M_{\text{S}}$ of the thin films is close
to the values measured in bulk samples.

For an ideal switching behavior of a TMR-device it is important
that both electrodes switch separately at different fields. Within
this field range, a stable antiparallel configuration of the
magnetization of the electrodes can be achieved. The difference in
resistance of the stable states with perfect parallel and
antiparallel order of the magnetization of the electrodes is a
key precondition for high TMR-effects \cite{Julliere:75}.
\begin{figure}[htp]
 \centering
  \includegraphics[width=0.9\columnwidth, trim=10 10 10 10]{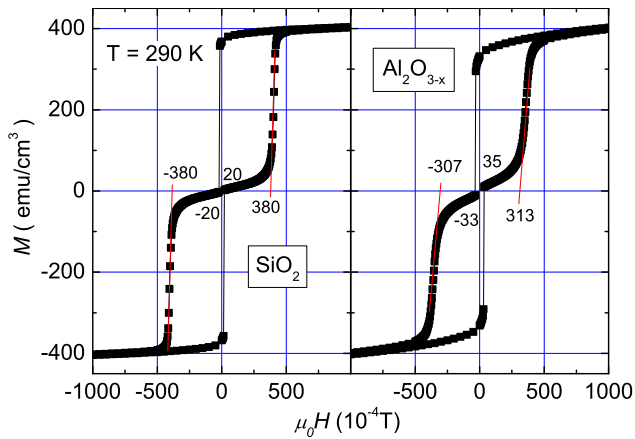}
 \caption{$M(H)$ at room temperature for two different barrier materials. Left panel: Fe$_3$O$_4$/SiO$_2$/Ni. Right panel: TiN/Fe$_3$O$_4$/Al$_2$O$_{3-x}$/Ni.}
 \label{fig3}
\end{figure}
\vspace{-0.2cm}

We first discuss the behavior of MgO barriers. As shown in
Fig.~\ref{fig2}, for a barrier of 5.3\,nm MgO the difference of
the switching fields of Fe$_3$O$_4$ and Ni is only about 7\,mT at
room temperature. At lower temperatures (210\,K and 150\,K) the
electrodes switch together around applied fields of 30\,mT. These
switching fields are far away from the typical coercive fields for
Ni thin films ($\approx7$\,mT at 5\,K [$M_{\text{S}}(5\,\text{K},
7\,\text{T})\approx560\,\text{emu/cm$^3$}$] and $\approx0.25$\,mT
at room temperature [$M_{\text{S}}(290\,\text{K},
7\,\text{T})\approx495\,\text{emu/cm$^3$}$]) which are much
smaller as observed in the multilayer structure. It is evident
that the electrodes are strongly coupled through the MgO barrier
layer, behaving like almost one ferromagnetic layer.

To investigate the magnetic coupling through the MgO tunnel
barrier in more detail, the same measurements have been repeated
with a thicker barrier layer (21\,nm). From atomic force
microscopy and x-ray reflectometry measurements indicating a
surface roughness below one nm (root mean square), we exclude that
short circuits exist through the barrier. Interesting enough, the
same coupling phenomenon is observed again. At this moment, we can
only speculate about the origin of this phenomenon. The first
possibility is that Fe diffuses into the MgO barrier and vice
versa leading to a diluted magnetic system Mg$_{1-x}$Fe$_x$O.
While such interdiffusion has only been observed after annealing
processes at much higher temperatures \cite{Shaw:00}, here the PLD
process with a cloud of $\approx50$\,eV particles could drive
interdiffusion between the two well lattice matched systems MgO
and Fe$_3$O$_4$. In contrast, electron beam evaporation or plasma
assisted molecular beam epitaxy \cite{Shaw:00} deals with almost
thermalized atoms. Within this model, at increasing temperature
the coupling is decreased due to the decreased order of the
magnetic insulating layer. The second possible origin of the
magnetic coupling through the MgO barrier, is related to recent
ideas about new types of ferromagnets due to the presence of
specific point defects in low concentration \cite{Elfimov:02}.
Point defects due to Mg vacancies within such models are
responsible for \glq local\grq\ magnetic moments. Also, oxygen
deficient systems might in a similar way lead to ferromagnetic
behavior. However, for the purpose of TMR devices, non-magnetic
insulating tunnel barriers are needed that decouple the electrodes
effectively.

Fig.~\ref{fig3} shows magnetization measurements of two samples
consisting of 47~nm Fe$_3$O$_4$~/ 5~nm SiO$_2$~/ 40~nm Ni
resp.~14~nm TiN~/ 49~nm Fe$_3$O$_4$~/ 2.5~nm Al$_2$O$_{3-x}$~/
32~nm Ni. In contrast to the samples with MgO tunnel barrier
layer, these samples show a clear separated magnetization
switching of the electrodes at all temperatures. The coercive
field of the Ni layer is slightly increased as compared to the
values of the single thin reference nickel film. This effect is
due to residual coupling through the thin insulating barrier. For
the Al$_2$O$_{3-x}$ barrier the switching steps are more rounded
than for the case of SiO$_2$ indicating that magnetic domains do
not switch simultaneously within the whole interface region. The
difference between SiO$_2$ and Al$_2$O$_{3-x}$ is that SiO$_2$ and
Ni is evaporated directly on top of magnetite. In the case of
Al$_2$O$_{3-x}$, first an Al layer is deposited which is then
oxidized. It is this oxidation step that also can affect the
magnetite top layer leading to the observed non-ideal switching
behavior.

\subsection{$\mathbfit{R(T)}$}

A first crucial test for the reproducibility of the tunnel barrier
is the investigation of the product of junction area $A$ and
junction resistance. We have listed in Table~\ref{mtmr44-RF} the
measured room temperature values for a sample consisting of four
contacts with different size but with the same $2.5$\,nm
Al$_2$O$_{3-x}$ barrier. The four-point-resistance $R_{4p}$ of the
contacts scales inverse to their area, and the
resistance-times-area-product $R_{4P}\cdot A$ of all contacts is
in the same range of $10^{-9}\,\Omega \mathrm{m}^2$. This shows,
that all contacts are approximately of the same quality,
independent on their size. This implies that the insulating
SiO$_2$ coating at the border of the TMR-turrets indeed suppresses
leakage currents. The values obtained here for Al$_2$O$_{3-x}$
barriers are comparable to the results of other groups. Chen {\em
et al.} found 1-230$\cdot 10^{-9}\,\Omega$m$^2$ at room
temperature for simple oxidized barriers and four times higher
values with plasma oxidation \cite{Chen:00}. The junctions with
the higher resistance times area product also had a significantly
higher TMR effect \cite{Chen:00}. This shows that an optimization
of the Al oxidation process is a crucial step in the fabrication
of a TMR device.

\begin{table}
\centering
\begin{tabular}{|c|c|c|c|}
    \hline
    area $A$  &  $2p$-resistance  & $4p$-resistance & $R_{4p}\cdot A$\\ \hline
    10$\mu$m~x~10$\mu$m & 172~$\Omega$ & 66~$\Omega$ & 6.6$\cdot 10^{-9}\, \Omega$m$^2$ \\ \hline
    10$\mu$m~x~20$\mu$m & 292~$\Omega$ & 36~$\Omega$ & 7.1$\cdot 10^{-9}\, \Omega$m$^2$ \\ \hline
    20$\mu$m~x~20$\mu$m & 407~$\Omega$ & 23~$\Omega$ & 9.1$\cdot 10^{-9}\, \Omega$m$^2$ \\ \hline
    20$\mu$m~x~40$\mu$m & 539~$\Omega$ & 5.0~$\Omega$ & 4.0$\cdot 10^{-9}\, \Omega$m$^2$ \\ \hline
\end{tabular}
\caption{Resistance-times-area-product of a TMR-samples having a
2.5~nm Al$_2$O$_{3-x}$ barrier at room temperature.}
\label{mtmr44-RF}
\end{table}

\begin{figure}[b]
 \centering
  \includegraphics[width=0.9\columnwidth, trim=10 10 10 10]{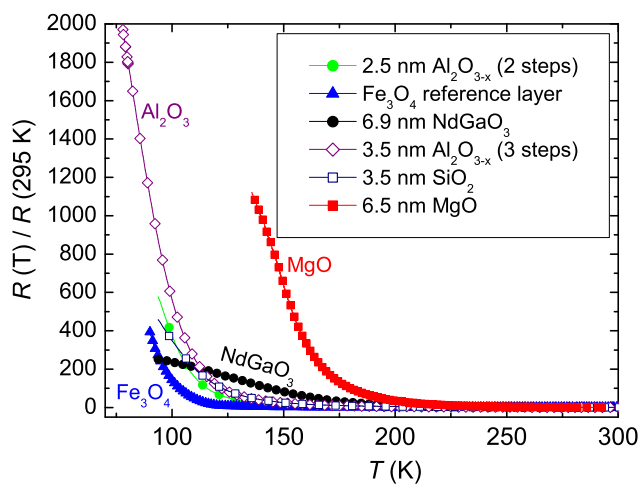}
 \caption{$R(T)$ of TMR-contacts with different tunnel barriers in comparison.}
 \label{rt-vergleich2}
\end{figure}

Fig.~\ref{rt-vergleich2} shows resistance versus temperature
curves of TMR-contacts with different tunnel barriers and a
magnetite reference film for comparison. All curves are normalized
to their room temperature resistance. The absolute resistance
values are in the range of 5\,$\Omega$~-~10\,k$\Omega$ at room
temperature. Due to the Verwey transition the resistance of the
magnetite layer increases several orders of magnitude at
approximately $T_{\text{V}}\approx120$\,K \cite{Verwey:39}. For
the multilayer systems the resistance increases already at
temperatures above $T_{\text{V}}$. This effect is most pronounced
for the MgO barrier layer where the resistance starts to increase
strongly at around 175\,K. As discussed above, it is likely that
especially the MgO barrier has a high density of dislocation
centers. Therefore, the increase of resistance is most likely due
to the suppression of thermally activated hopping conduction
through the insulating barrier.

Here, we briefly estimate the influence of the magnetite electrode
on the resistance of the whole turret structure. Assuming
homogeneous current feed into the turret due to the well
conducting TiN buffer layer, and knowing the resistivity of
magnetite, a resistance of 5.6$\cdot 10^{-3}\,\Omega$ at room
temperature is expected for a 20~x~20~$\mu$m$^2$ turret. It is
evident that the effect of Fe$_3$O$_4$ is negligible at least at
temperatures not to close to the Verwey transition. In the case of
inhomogeneous current feed, the influence of the magnetite
electrode can no longer be neglected. It is therefore necessary
for a precise determination of the barrier properties to have a
low resistance buffer layer beneath the Fe$_3$O$_4$ electrode.

\subsection{$\mathbfit{U(I)}$}

To further investigate the tunnel barrier properties,
$U(I)$-curves were measured and fitted to the Simmons model
assuming a trapezoid barrier shape \cite{Simmons:63}. From this
fit, the effective barrier height $\phi$ and barrier width $d$ can
be determined in the case $eV\ll\phi$. Due to the large barrier
heights $\phi$ in the $eV$-range, the measurement current has to
be chosen high enough in order to observe non-linearities. For
samples with low resistance of only few $\Omega$ the application
of large currents is of course not possible. The non-linearity in
$U(I)$ is most pronounced at low temperatures ($\approx 160$\,K),
and is replaced by almost Ohmic behavior at room temperature due
to increased thermally activated transport via defect states.
\begin{table}
\centering
\begin{tabular}{|c|c|c|c|c|}
    \hline
    barrier         & thickness $d$ & fit value $d$ & $\phi$ at 160\,K  & $\phi$ at 295\,K \\ \hline
    NdGaO$_3$       & 6.9\,nm       & 0.9\,nm       & 5\,eV             & 2\,eV \\ \hline
    SrTiO$_3$       & 2.8\,nm       & 1.4\,nm       & 1.6\,eV           & 1.5\,eV \\ \hline
    SiO$_2$         & 3.5\,nm       & 1.4\,nm       & 1.2\,eV           & 0.9\,eV \\ \hline
    MgO             & 6.5\,nm       & 2.3\,nm       & 0.9\,eV           & 0.7\,eV \\ \hline
    Al$_2$O$_{3-x}$ & 2.5\,nm       & 1.6\,nm       & 0.7\,eV           & 0.6\,eV \\ \hline
\end{tabular}
\caption{Effective barrier height $\phi$ and width $d$ (at 160\,K) of the tunnel barrier in
samples with different barrier materials as derived using the
Simmons model \cite{Simmons:63}.} \label{Barrierentab1}
\end{table}
Using the Simmons model the $U(I)$ curves could be fitted very
satisfactory. The results for the effective barrier height $\phi$
and width $d$ after the Simmons-model for the different tested
barrier materials are listed in Table~\ref{Barrierentab1}.
NdGaO$_3$ with 5\,eV has the highest barrier, and Al$_2$O$_{3-x}$
with 0.7~eV the lowest barrier height at the Fermi level. These
values are well comparable to the results of other groups, see for
example $\phi=0.9$\,eV for MgO \cite{Kiyomura:00}. The value for
the effective barrier width as obtained from the Simmons fit is in
most cases considerably smaller than the nominal barrier
thickness. It becomes clear from Table~\ref{Barrierentab1} that
for Al$_2$O$_{3-x}$ barriers the fit parameter $d$ is closest to
the nominal barrier thickness indicating that Al$_2$O$_{3-x}$
seems to be the best material for obtaining thin and at the same
time well insulating barriers. For comparison, in order to achieve
comparable barrier properties using MgO a several times thicker
layer has to be used. It can be assumed that the increased barrier
thickness leads to increased spin scattering reducing the TMR
effect. This effect is indeed observed as shown below. The reason
that Al$_2$O$_{3-x}$ is the most favorable barrier is most likely
due to the fact that during electron beam evaporation first an
almost complete surface wetting is achieved with elementary Al.

\begin{figure}[b]
 \centering
  \includegraphics[width=0.9\columnwidth, trim=10 10 10 10]{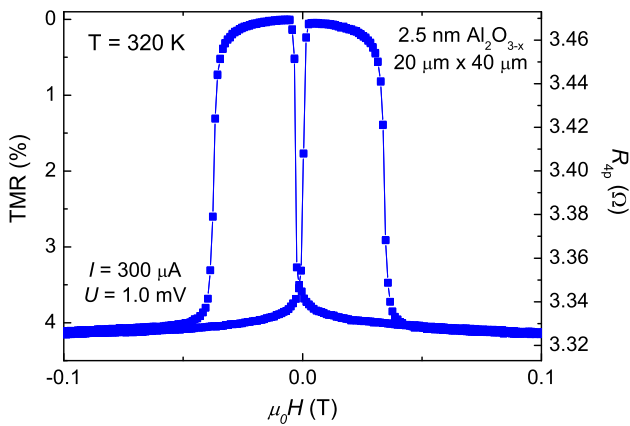}
 \caption{TMR-effect at room temperature of a sample with 2.5\,nm Al$_2$O$_{3-x}$
 barrier. The barrier was oxidized in a 1-step process.}
 \label{Alu-TMR}
\end{figure}

\subsection{TMR-effect}

With $R(H)$-measurements in 4-point-technique the TMR-effect as
defined in equation~\ref{tmreqdef} was determined for the
different barrier materials at different temperatures between
160\,K and 300\,K:
\begin{equation}\label{tmreqdef}
TMR=\frac{R_{ap}-R(H)}{R_p}.
\end{equation}
$R_{ap}$ is the resistance in the anti-parallel magnetization
configuration of the electrodes, which corresponds to the maximum
resistance in figure \ref{Alu-TMR}. $R_p$ is the resistance in the
parallel magnetization configuration of the electrodes, which
corresponds to the minimal resistance at 0.1 Tesla.

The samples with MgO and SiO$_2$ barrier exhibited only a small
positive TMR-effect $<$~0.5\% at low temperatures. Samples with
SrTiO$_3$ and NdGaO$_3$ barrier showed nearly no TMR-effect at
all. In these multilayers a small negative conventional
magnetoresistance was observed as is the case for magnetite thin
films. This indicates that the electrode properties dominate the
measurements in these contacts. It has been shown by other groups
\cite{DeTeresa:99,Bowen:03} that it is possible to obtain
TMR-behavior using these barrier materials at low temperatures.
But in the case of magnetite electrodes however, it is not easy to
find good TMR-behavior below $T_{\text{V}}$ due to the strong
increase of resistance.

The samples with Al$_2$O$_{3-x}$ barrier showed reproducibly a
clear positive TMR-effect with almost ideal, symmetric switching
behavior as shown in Fig.~\ref{Alu-TMR}. The TMR effect is
observable in the whole measured temperature range from 150\,K to
350\,K. At room temperature up to 5\% resistance change was found.
A second oxidation process in our case did not increase the TMR
effect. The TMR-behavior is in agreement with the analysis of the
junction properties. Al$_2$O$_{3-x}$ provides a good insulator for
the thinnest barriers with the lowest amount of defect states
leading to magnetic decoupled electrodes and reduced spin
scattering at the same time. Note that the increase of resistance
for antiparallel electrode magnetization (positive TMR-effect)
confirms the negative spin-polarisation of Fe$_3$O$_4$. Since Ni
and Fe$_3$O$_4$ both have negative spin polarisation, the observed
behavior is expected. For an electrode with positive spin
polarisation as the double exchange material
La$_{0.7}$Sr$_{0.3}$MnO$_3$, a negative TMR-effect is expected,
and indeed observed \cite{DeTeresa:99b,Hu:03}.

\begin{figure}[t]
 \centering
  \includegraphics[width=0.9\columnwidth, trim=10 10 10 10]{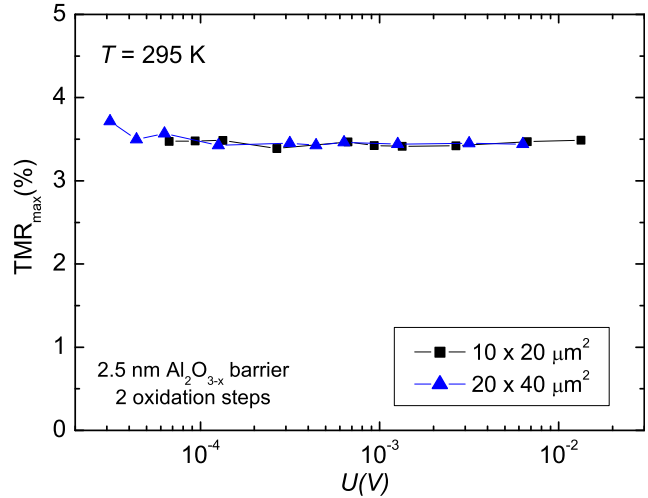}
 \caption{TMR-effect versus voltage for samples with 2.5~nm Al$_2$O$_{3-x}$ barrier at room temperature.}
 \label{tmr-vs-U}
\end{figure}

It is well known that for samples with a high density of defect
states, at increased voltages the TMR effect decreases due to
inelastic spin scattering transport processes in the barrier
\cite{Hoefener:00}. In Fig.~\ref{tmr-vs-U} the maximal TMR-effect
is plotted versus the applied tunnel voltage. The different
measurement points were obtained for different applied currents.
It is clear that over more than two orders of magnitude (in the
range of $10^{-5}$~V to $10^{-2}$~V) no change of the TMR effect
is observed. This result implies that in the observed voltage
range no additional inelastic transport channels with spin
scattering are opened. The Al$_2$O$_{3-x}$ barrier acts as a good
insulating barrier where the defect states do not dominate the
transport behavior in the low-voltage regime.

\begin{figure}[floatfix]
 \centering
  \includegraphics[width=0.95\columnwidth, trim=10 20 10 20]{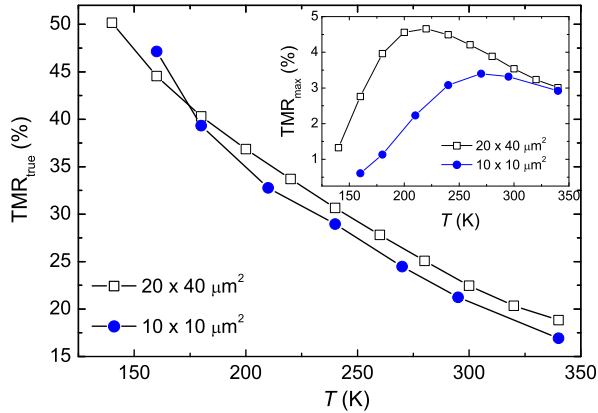}
 \caption{\glq True\grq\ TMR effect and maximal TMR effect versus temperature for
           samples with 2.5\,nm Al$_2$O$_{3-x}$ barrier.}
 \label{tmr-vs-t}
\end{figure}

The inset of Fig.~\ref{tmr-vs-t} shows the temperature dependence
of the maximal TMR-effect $TMR_{\text{max}}$ of two samples with
2.5~nm Al$_2$O$_{3-x}$ barrier. The junction size and the number
of oxidation steps varies for both junction. The larger junction
($20\times40\,\mu$m$^2$) has a two-step oxidized barrier
\cite{2step}, while the smaller junction ($10\times10\,\mu$m$^2$)
has a one-step oxidized barrier. $TMR_{\text{max}}(T)$ has a
non-linear behavior with a maximum the position of which depends
on deposition details. For higher temperatures above approximately
280\,K all curves converge. It is obvious that only a decrease of
magnetization/spin polarisation in the tunneling electrodes and
thermally assisted inelastic spin scattering processes lead to a
decreased TMR effect. $TMR_{\text{max}}(T)$ is expected to be a
continously decreasing function with increasing temperature. The
reason that experimentally a maximum is observed in
$TMR_{\text{meas}}(T)$ can be explained within a simple model
assuming a constant series resistance $R_{S}$ in the TMR-turret in
addition to the barrier resistance $R_B$ . The series resistance
is not affected by the change of spin polarisation of the
electrodes producing the \glq real\grq\ TMR-effect by switching
$R_B$. For the measured TMR effect $TMR_{\text{meas}}$ one
obtains:
\begin{equation}
TMR_{\text{meas}}=\frac{\Delta R}{R_B + R_S}
 \label{tmrtrue1}
\end{equation}
This value can easily be related to the \glq true\grq\ TMR effect
$TMR_{\text{true}}$:
\begin{equation}
TMR_{\text{true}}=\frac{\Delta
R}{R_B}=TMR_{\text{meas}}\left(1+\frac{R_S}{R_B}\right)
 \label{tmrtrue2}
\end{equation}
Because the measured 4-point-resistance $R_{\text{4p}}$ of the
TMR-structure is the sum of the barrier resistance $R_B$ and the
series-resistance $R_S$, eq.~\ref{tmrtrue2} can be rewritten as
\begin{equation}
TMR_{\text{true}}=\frac{\Delta
R}{R_B}=TMR_{\text{meas}}\left(1+\frac{R_{\text{4p}}-R_B}{R_B}\right).
 \label{tmrtrue3}
\end{equation}
While $R_{\text{4p}}$ is a measured quantity, the value of $R_B$
has to be estimated. By assuming a realistic resistance times area
product of approximately $10^{-9}\,\Omega \mathrm{m}^2$, for $R_B$
one obtains roughly 10\,$\Omega$. The resulting
$TMR_{\text{true}}(T)$ is shown in Fig.~\ref{tmr-vs-t}, and has
the expected continuously (almost linearly) decreasing behavior.
Between 150\,K and 300\,K $TMR_{\text{true}}$ decreases by more
than 50\% while the corresponding decrease in magnetization is
only about 10\% for Fe$_3$O$_4$ and 11.6\% for Ni. According to
the Julli\`{e}re model \cite{Julliere:75}, this accounts for a
decrease in $TMR_{\text{true}}$ of 28\% assuming that
magnetization translates directly into spin polarisation. Having
in mind the results of $U(I)$ at different temperatures and also
the reduced magnetic coupling, the difference to the estimated
value of $TMR_{\text{true}}$ can be understood due to the increase
in spin-scattering at defect states in the tunnel-barrier.

\section{Summary}
We have successfully prepared magnetic tunnel junctions with sizes
of $10\times 10\,\mu$m$^2$ to $20\times 40\,\mu$m$^2$ with optical
lithography and ion beam etching. The bottom electrodes consist of
epitaxial magnetite thin films on MgO substrates using a TiN
buffer-layer. As tunnel barrier five different materials have been
investigated. Thin nickel films have been used for the top
electrode.

The analysis of $M(H)$ curves showed that for MgO barriers the
electrodes couple magnetically even for high barrier thicknesses
(up to 21\,nm). Al$_2$O$_{3-x}$ as barrier leads to a clearly
separated magnetization switching of the electrodes with more
rounded steps than in the case of SiO$_2$.

$R(T)$- and $U(I)$-measurements of the tunnel-contacts showed a
strong decrease of the insulating behavior of the barriers with
increasing temperature. This may be due to tunneling via thermal
activated defect states in the barrier layer.

The TMR-contacts with MgO and SiO$_2$ barrier exhibited only a
small effect $<$\,0.5\% at low temperatures $< 200$\,K. Samples
with SrTiO$_3$ and NdGaO$_3$ barrier showed nearly no TMR-effect
at all. The tunnel-contacts with Al$_2$O$_{3-x}$ barrier showed
reproducibly a clear TMR-effect with ideal, symmetric switching
behavior. The effect was visible in the whole investigated
temperature range from 150\,K to 350\,K. At room temperature up to
5\% resistance change was found. Almost no influence of the
applied voltage on the size of the TMR-effect in the studied
voltage range from $10^{-5}$\,V to $10^{-2}$\,V is observed. The
temperature dependence of the as measured TMR-effect shows a
pronounced maximum. This can be explained by an additional
series-resistor in the TMR-turret combined with the decrease of
the spin-polarization of the electrodes at increasing temperature,
and by increased spin-flip-scattering at higher temperatures.
Further experiments are needed to specify the exact value of the
spin-polarization of the magnetite electrode at room temperature.
But it is obvious from the presented data, that magnetite has high
spin-polarization and certainly is an interesting candidate as a
material for further spinelectronic devices at room temperature.

We would like to thank J.~Schuler for help with the electron beam
evaporation and the lithography. This work was supported in part
by the Deutsche Forschungsgemeinschaft (project No.:\,Al/560) and
the BMBF (project No.:\,13N8279).



\begin{thebibliography}{10}
\bibitem{2step}
The barrier is deposited in two steps, and after each deposition
step oxidation is performed.
\end{thebibliography}
\end{document}